\documentclass[aps, prl, twocolumn, superscriptaddress, amsmath, tightenlines, longbibliography]{revtex4-1}
\usepackage[none]{hyphenat}
\usepackage{amsmath}
\usepackage{amssymb}
\usepackage{amsfonts}
\usepackage{physics}
\usepackage{graphicx}
\usepackage{epsfig}
\usepackage{color}
\usepackage{textcomp}
\usepackage[colorlinks,citecolor=blue]{hyperref}
\usepackage{mathrsfs}
\usepackage{appendix}
\usepackage{dcolumn}
\usepackage{booktabs}
\usepackage{units}
\usepackage{url}
\usepackage[colorlinks]{hyperref}
\hypersetup{%
	plainpages=true,
	breaklinks=true,       %not default in dvips mode, so we must specify
	hypertexnames=false,  %not ideal, but needed when pagenums duplicate (`i' vs. `1')
	pageanchor=true,
	colorlinks=true,
	linkcolor={blue},
	citecolor={blue},
	urlcolor={blue},
	anchorcolor={black}
}
\begin{document}
\title{Nonreciprocal Generation of Schr\"{o}dinger Cat State Induced by Topology}
\author{Zi-Hao Li}
\affiliation{School of Physics and Institute for Quantum Science and Engineering, Huazhong University of Science and Technology, Wuhan 430074, P. R. China}

\author{Li-Li Zheng}
\affiliation{Key Laboratory of Optoelectronic Chemical Materials and Devices of Ministry of Education, Jianghan University, Wuhan 430074, China}

\author{Ying Wu}
\affiliation{School of Physics and Institute for Quantum Science and Engineering, Huazhong University of Science and Technology, Wuhan 430074, P. R. China}

\author{Xin-You L\"{u}}\email{xinyoulu@hust.edu.cn}
\affiliation{School of Physics and Institute for Quantum Science and Engineering, Huazhong University of Science and Technology, Wuhan 430074, P. R. China}

\date{\today}% It is always \today, today,
\begin{abstract}
The Schr\"{o}dinger cat state produced differently in two directions is anticipated to be a critical quantum resource in quantum information technologies. By exploring the interplay between quantum nonreciprocity and topology in a one-dimensional microcavity array, we obtain the Schr\"{o}dinger cat state ({\it a pure quantum state}) in a chosen direction at the edge cavity, whereas a {\it classical state} in the other direction. This {\it nonreciprocal generation of the cat state} originates from the {\it topologically protected chirality-mode excitation} in the nontrivial phase, but in the trivial phase the {\it nonreciprocal generation of cat state} vanishes. Thus, our proposal is switchable by tuning the parameters so that a topological phase transition occurs. Moreover, the obtained cat state has nonreciprocal high fidelity, nonclassicality, and quantum coherence, which are sufficient to be used in various one-way quantum technologies, e.g., invisible quantum sensing, noise-tolerant quantum computing, and chiral quantum networks. Our work provides a general approach to control quantum nonreciprocities with the topological effect, which substantially broadens the fields of nonreciprocal photonics and topological physics.
\end{abstract}
%corresponding excitations

\maketitle
The Schr\"{o}dinger cat state, as a macroscopic quantum superposition state, not only plays a fundamental role in exploring the quantum-to-classical transition, but also is crucial to quantum metrology\,\cite{Kira2011, Joo2011} and fault-tolerant quantum computation\,\cite{Ralph2003, Mirrahimi2014}. The normal cat states can be obtained by two approaches: (i) non-Gaussian measurement, which typically using the technique called photon subtraction\,\cite{Nielsen2006, Sun2021}, and (ii) the nonlinear effect, e.g., single-photon Kerr effect\,\cite{Kirchmair2013}, optomechanical interaction\,\cite{Bose1997}, and so on\,\cite{Agarwal1997, Munoz2018, Qin2021}. However, in previous schemes, these cat states are not prepared to have difference in directions, thus cannot be used to apply one-way quantum signal propagation, chiral quantum networks and invisible quantum sensing. This suggests a strong demand of {\it nonreciprocal generation of Schr\"{o}dinger cat states} for the development of unidirectional quantum techniques\,\cite{Lodahl2017}. The previous studies of optical nonreciprocities were primarily focus on the {\it nonreciprocal classical effects} (e.g., the transmission of light), and various methods have been proposed based on the spatial and temporal modulation of material\,\cite{Yu2009, Sounas2017}, parity-time symmetry breaking\,\cite{Peng2014,Chang2014}, etc.\cite{Hafezi2012, Shen2016, Maayani2018, Wang2019, Liang2020, Tang2022, 2_Tang2022}. By contrast, research on {\it quantum nonreciprocity} is still in the ascendant. Recently, by utilizing the existing techniques for achieving classical nonreciprocity, purely quantum nonreciprocal effects such as single-photon isolation\,\cite{Dong2021}, nonreciprocal photon, phonon or magnon blockade\,\cite{Huang2018,Li2019, Yao2022, Wang2022}, and nonreciprocal entanglement\,\cite{Jiao2020, Yang2020, Jiao2022, Liu2023} have been investigated. Moreover, the nonreciprocal quantum correlations of photons can be achieved even when no classical nonreciprocity exists\,\cite{Graf2022, Xiang2023}. This suggests that the realization condition for quantum nonreciprocity might be different from the condition for classical nonreciprocity. In particular, the {\it nonreciprocal generation of the Schr\"{o}dinger cat state} has not been proposed in any system until now, since the previous proposals for controlling classical nonreciprocities cannot be easily used to realize the nonreciprocal generation of quantum states.
\begin{figure}[t]
  \centering
  \includegraphics[width=8.5cm]{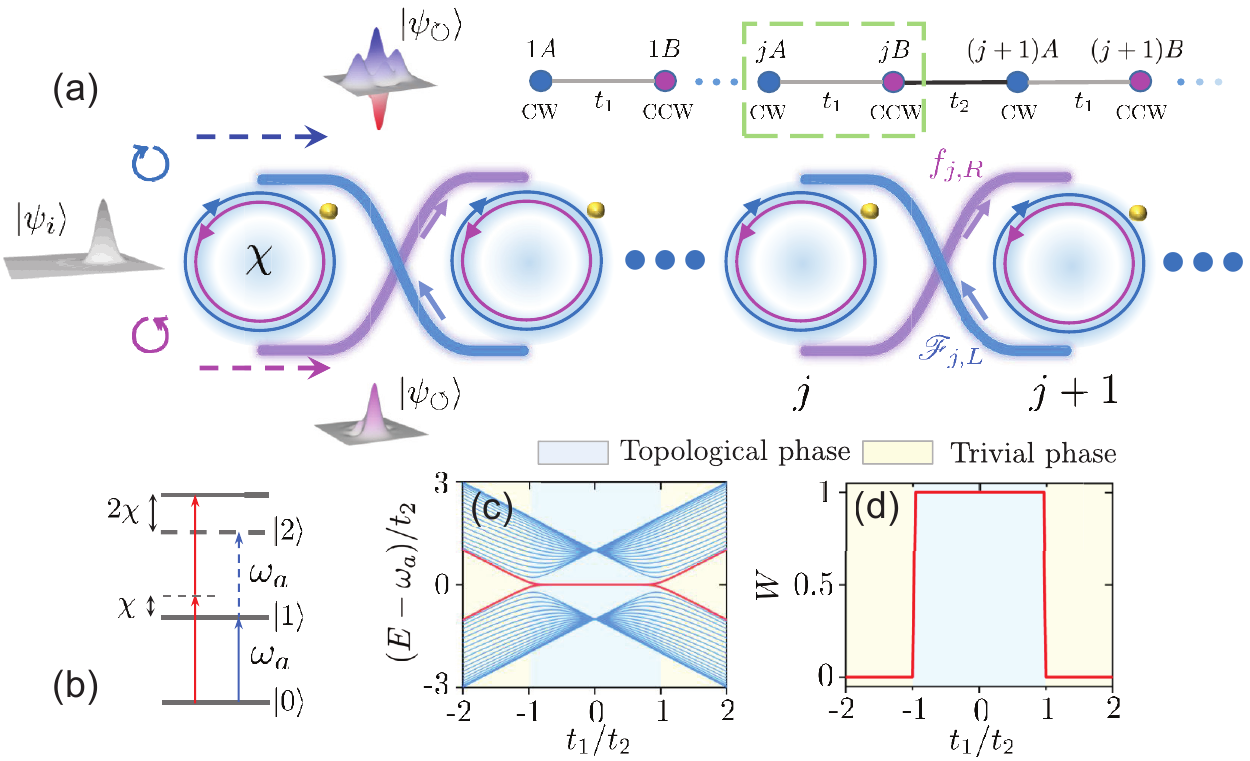}
  \caption{(Color online) (a) Scheme of nonreciprocal generation of the Schr\"{o}dinger cat states consists of a 1D microtoroid cavity array. Each cavity supports two counter-circulating modes that are coupled with each other via a scatterer, and the adjacent cavities are connected by the unidirectional fibers. The cavity index number is denoted by $j$ below the cavity, $f_{j,\rm R}$ and $\mathscr{F}_{\rm L}$ are the right propagating mode in the $j$th purple fiber and the left propagating mode in the $j$th blue fiber, respectively. The left-edge cavity in this array is a Kerr cavity with nonlinearity $\chi$. The cavity array with zero nonlinearity $ \chi=0 $ can be described by an SSH model shown at the right top. The dashed box is a unit cell. $A$ and $B$ are two sublattices that assigned to the CW and CCW modes, and $t_{1}$, $t_{2}$ refer to the on-site and off-site coupling strengths, respectively. Given the same initial state $|\psi_i \rangle$ in the two modes of the left-edge cavity, a high-fidelity Schr\"{o}dinger cat state $|\psi_{\circlearrowright} \rangle$ can be prepared in the CW mode, whereas a classical state $|\psi_{\circlearrowleft} \rangle$ can be obtained in the opposite direction. (b) Energy level of the edge cavity is shifted by the Kerr nonlinearity. The blue and red solid arrow lines refer to the single-photon and two-photon excitations, respectively. (c,d) The energy spectral of an open array with 20 cavities and the corresponding winding number versus $t_{1}/t_{2}$. The red lines in (c) mark the edge states.}\label{fig1}
\end{figure}

The topological concept, originating from the study of solid-state electron systems, has been extended to bosonic systems\,\cite{Haldane2008, Lu2014, Peano2015, Huber2016, Tuo2019, Ozawa2019, Chen2019, Gao2020} and generates substantial exotic quantum optics effects, including topologically protected quantum coherence~\cite{2Wang2019, Nie2020}, topological lasing~\cite{Harari2018,Bandres2018}, and so on.~\cite{Malzard2015, Bernardis2021}. In two-dimensional (2D) topological systems, breaking time-reversal symmetry with a synthetic magnetic field can result in the occurrence of topologically protected one-dimensional (1D) chiral edge states propagating along multiple boundary sites\,\cite{Wang2008, Goldman2012}. These chiral edge states are robust against disorder and immune to backscattering, thus have been widely implemented in the optical information processing\,\cite{Hafezi2011, Dlaska2017}, especially in the realization of classical optical nonreciprocities, e.g., optical isolator and unidirectional light propagation\,\cite{Mittal2014, Bahari2017, Solnyshkov2018}. However, they have difficulty in forming the nonreciprocal generation of single-site quantum states owing to their nonlocality. Moreover, the nonreciprocal zero-dimensional (0D) topological edge state is still not observed. In photonic systems such as microtoroid cavity arrays, the bidirectional 0D edge naturally exists(the clockwise (CW) and counterclockwise (CCW) traveling modes in one single cavity). Therefore, the topological photonic systems provide a probability for finding the nonreciprocal 0D topological edge state and then become a prospective platform for {\it nonreciprocal generation of the Schr\"{o}dinger cat state}.

Inspired by the above discussion, we present a scheme to realize {\it the topologically induced nonreciprocal generation of the Schr\"{o}dinger cat state} in a 1D microtoroid cavity array. This 1D system is a photonic analog of the Su-Schrieffer-Heeger (SSH) model, and there exists a topological phase transition when moving the nanoparticles placed beside the cavities. Interestingly, the {\it nonreciprocal generation of the Schr\"{o}dinger cat state} is achieved in the edge Kerr cavity when the 1D array is in the topological nontrivial phase. Specifically, a high-fidelity Schr\"{o}dinger cat state ({\it quantum state}) can be prepared in the CW mode, but only a classical state can be prepared in the CCW mode with the same initial condition. Physically, the {\it topologically protected chirality-mode excitation} performed in two counter-propagating modes of the edge cavity induces the appearance of the nonreciprocal Kerr effect, which ultimately leads to the nonreciprocal generation of the Schr\"{o}dinger cat state. In particular, the nonreciprocal single-photon band structure in an analog SSH coupled-cavity array has been proposed very recently~\cite{Tang2022}. Different from our topology-induced nonreciprocity, the nonreciprocal band structure in this work is induced by the chiral interaction between the resonator array and quantum emitters.

Moreover, we also show that the key properties of the obtained cat state, including Wigner negativity, macroscopic quantum superposition, and fidelity~\cite{Sun2021}, also suggest strong nonreciprocity of cat state generation in the topological nontrivial phase. This quantum nonreciprocity vanishes in the trivial phase, which offers the important family of topological phase transitions with a new type of application in controlling nonreciprocal quantum resources. As far as we know, the {\it nonreciprocal generation of the Schr\"{o}dinger cat state} is proposed for the first time, which is not only fundamentally interesting for quantum physics, but also has wide applications ranging from one-way quantum information processing to constructing chiral quantum networks. Our proposed topological-induced nonreciprocal mechanism in our proposal is general, and can be extended to realize nonreciprocal generation of other quantum states, e.g., Fock state and NOON states.

\emph{System and nonreciprocal Kerr effect.}---We consider a 1D microtoroid cavity array shown in Fig.\,\ref{fig1}(a), where every toroid cavity supports the CW and CCW traveling modes with the same resonance frequency $\omega_a$. A nanoparticle is placed in the mode volume of each cavity, and can induce a coupling between the CW and CCW modes with strength $t_1$ by Rayleigh scattering. This coupling can be controlled by manipulating the position of scatterer with a nanopositioner\,\cite{Zhu2010, Peng2016}. The neighboring cavities are connected by two unidirectional fibers\,\cite{Yu2009}, which enables a bidirectional coupling between the CCW mode of the left cavity and the CW mode of the right cavity. The edge cavity is a Kerr resonator, then the on-site Hamiltonian is
\begin{align}
	H_{\rm c}=&\sum^{N}_{j=1}[\omega_a (a_{j,\circlearrowright}^{\dagger}a_{j,\circlearrowright}+ a_{j,\circlearrowleft}^{\dagger}a_{j,\circlearrowleft})\!+\!(t_{1} a_{j,\circlearrowright}^{\dagger}a_{j,\circlearrowleft}+h.c.)]\nonumber
	\\
	&+\chi (a_{1,\circlearrowright}^{\dagger}a_{1,\circlearrowright}^{\dagger}a_{1,\circlearrowright}a_{1,\circlearrowright}+a_{1,\circlearrowleft}^{\dagger}a_{1,\circlearrowleft}^{\dagger}a_{1,\circlearrowleft}a_{1,\circlearrowleft}),
\end{align}
where $a_{j,\circlearrowright} (a_{j,\circlearrowleft})$ is the annihilation operator of the CW (CCW) mode in the $j$th cavity. Here $\chi$ refers to the Kerr nonlinear susceptibility, which can be introduced by various methods such as Kerr material and electromagnetically induced transparency\,\cite{Schmidt1996,Brasch2016}. In Fig.\,\ref{fig1}(b), we present the energy structure of a single Kerr mode, with the Hamiltonian $ H_{\rm s} = \omega_a a^{\dagger}a + \chi a^{\dagger} a^{\dagger}a a $. Note that the first excited level is on-resonance, whereas the second excited level is detuned by $ 2\chi $ owing to Kerr nonlinearity. Corresponding to the two-photon excitation $| 0 \rangle \rightarrow | 2 \rangle$, the frequency of the driving photon is shifted by $\chi$.

The off-site link Hamiltonian can be written as
\begin{align}
	\!\!H_{l} =& i\sqrt{2\kappa_f}\sum^{N-1}_{j=1}[a_{j,\circlearrowleft}f^{\dagger}_{j,\rm R}(t,z_j)+a_{j,\circlearrowleft}\mathscr{F}^{\dagger}_{j,\rm L}(t,z_j)\nonumber
	\\&+a_{j+1,\circlearrowright}f^{\dagger}_{j,\rm R}(t,z_{j+1})+a_{j+1,\circlearrowright}\mathscr{F}^{\dagger}_{j,\rm L}(t,z_{j+1})\!-\!h.c.],
\end{align}
where $\kappa_f$ is the loss induced by the resonator-fiber coupling, $f_{j,\rm R}(t,z) = \frac{1}{\sqrt{2\pi}}\int_0^\infty d\omega f_{j,\omega} e^{- i\omega (t - z/c)} $ is the right propagating fiber mode in the $j$th purple fiber, and $\mathscr{F}_{\rm L}(t,z) = \frac{1}{\sqrt{2\pi}}\int_0^\infty d\omega \mathscr{F}_{j,\omega} e^{- i\omega (t + z/c)} $ is the left propagating fiber mode in the $j$th blue fiber. Here, the $j$th fiber is placed on the right of the $j$th cavity. Basically, the CCW mode of the $j$th cavity can construct a two-way photon transition with the CW mode of the ($j+1$)th cavity via the $j$th right and left propagating fibers. For the CW mode of the $j$th cavity, it can construct a two-way photon transition with the CCW mode of the ($j-1$)th cavity via the ($j-1$)th right and left propagating fibers. As we will discuss later in this section, this is important for building up a typical topological SSH structure in an arrayed toroid cavity, where the CW and CCW modes become different sublattices of the unit cell. We can obtain the quantum Langevin equations of every cavity modes, and then eliminate the fiber modes to get the effective couplings between cavities by using the Born-Markov approximation. The coupling coefficients are chosen to be real by modifying the length of fibers, and the effective link Hamiltonian thus becomes $\tilde{H}_{l} = \sum^{N-1}_{j=1} (t_{2} a_{j,\circlearrowleft}^{\dagger}a_{j+1,\circlearrowright}+t_{2} a_{j+1,\circlearrowright}^{\dagger}a_{j,\circlearrowleft})$, where $t_{2}$ is the coupling coefficient. In particular, the leftward hopping ($ a_{j,\circlearrowleft}^{\dagger}a_{j+1,\circlearrowright} $) arises from the blue fiber and the rightward hopping ($ a_{j+1,\circlearrowright}^{\dagger}a_{j,\circlearrowleft} $) arises from the purple fiber. As a result, the total Hamiltonian for the cavity array is $H_{\rm tot}=H_c+\tilde{H}_l$.

\begin{figure}
  \centering
  \includegraphics[width=8.5cm]{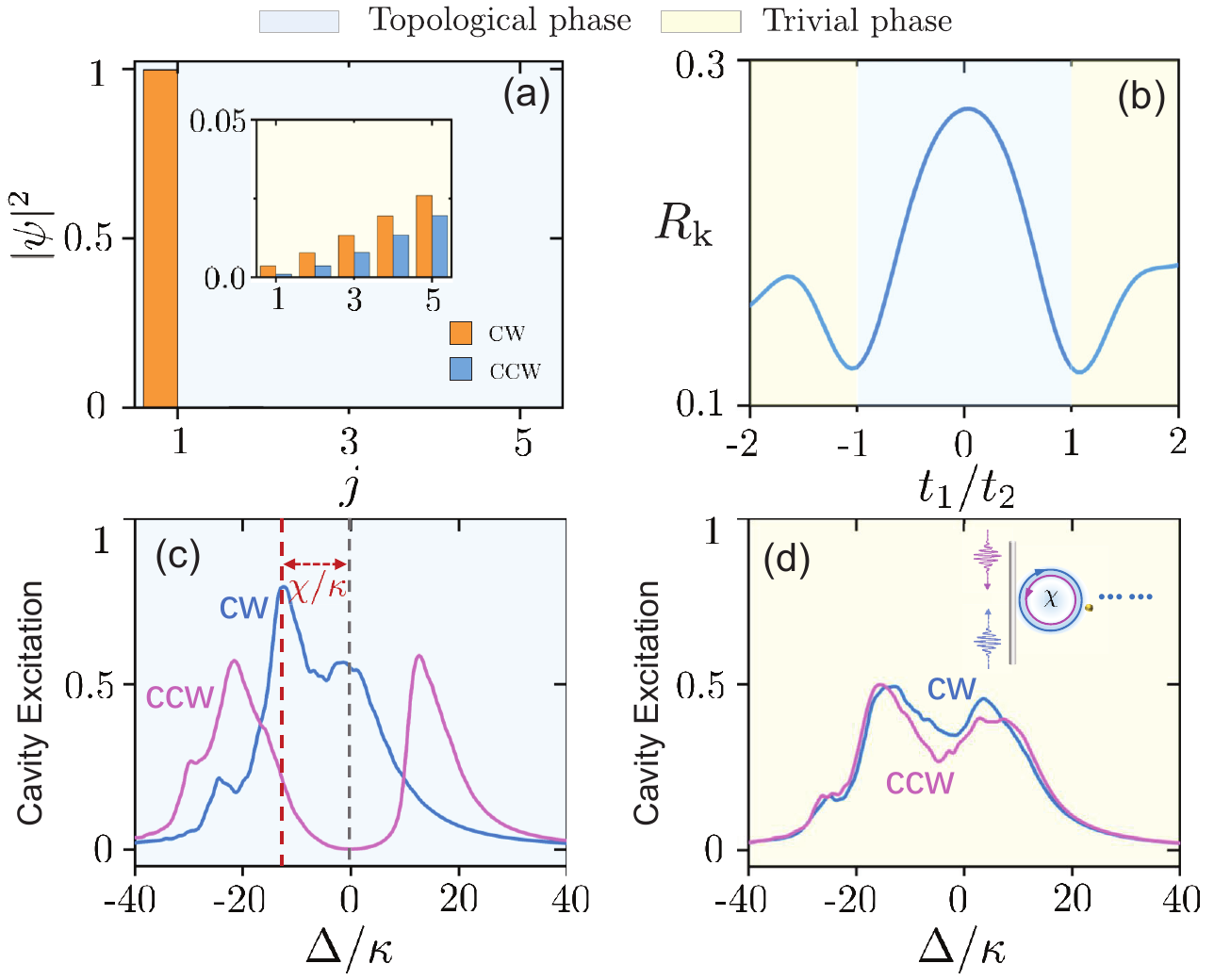}
  \caption{(Color online) (a) Profiles of the edge state in the topological nontrivial phase ($t_{1}/t_{2}=0.05$). The inset is the corresponding ground state in the trivial phase ($t_{1}/t_{2}=2$). (b) Difference rate of the chiral Kerr effect $R_{\rm k}$ versus $ t_{1}/t_{2} $ when $\chi=4\pi\kappa $ and $ t_{2} = 8\kappa$. (c,d) Excitation spectrum of the edge cavity with the present of the Kerr nonlinearity $\chi=4\pi\kappa $ in the (c) topological phase ($t_{1} = 0.8\kappa $, $ t_{2} = 16\kappa$) and (d) trivial phase ($t_{1} = 8\kappa $, $ t_{2} = 4\kappa$). The inset in (d) shows the edge cavity driven from opposite directions. Other parameters are $N=5$, $\gamma=\kappa$, and $\epsilon=8\kappa$.}\label{fig2}
\end{figure}
For zero Kerr nonlinearity, i.e., $\chi=0$, the total Hamiltonian $H_{\rm tot}$ describes an SSH model, where the CW and CCW modes serve as sublattices A and B, respectively. According to the topological insulator theory\,\cite{Asboth2016}, the system enters into the topological nontrivial phase when $t_{1}<t_{2}$, which associates with the degeneration of energy spectra and the occurrence of two edge states, as shown in Fig.\,\ref{fig1}(c). This process can also be carried by the winding number. Typically, we make Fourier transformation on cavity modes under the periodic boundary condition and obtain $a_{k,{\circlearrowright/\circlearrowleft}} = 1/\sqrt{N}\sum_j a_{j,{\circlearrowright/\circlearrowleft}}e^{ikj}$. Then we can write the Bloch Hamiltonian as $H_{\rm tot}(\chi=0) = \sum_{k} V^{\dagger} H_{\rm tot}(k) V$, where $V = (a_{k,{\circlearrowright}}, a_{k,{\circlearrowleft}})^{T}$ and $ H_{\rm tot}(k) = \omega_a I + d_x\sigma_x + d_y\sigma_y $ with $ d_x = t_1+t_2 \cos(k) $, $ d_y = t_2 \sin(k)$. Finally, the winding number is defined as
\begin{align}
	W = 1/(2\pi i)\int_{-\pi}^{\pi} dk \frac{d\,{\rm ln}h(k)}{dk},
\end{align}
where $h(k) = d_x-i d_y$, and the results are plotted in Fig.\,\ref{fig1}(d). To explore the property of edge states, we plot the profiles of one edge state in the topological phase and the corresponding ground state in the trivial phase in Fig.\,\ref{fig2}(a). Interestingly, in the topological phase, the CW mode of the first cavity is occupied and the corresponding CCW mode is topologically inhibited. In contrast, both the CW and CCW modes of the corresponding ground state can be occupied when the system enters into the trivial phase by adiabatically changing $t_1/t_2$. This leads to a {\it topologically protected chirality-mode-excitation} in the single edge cavity when the microtoroid cavity array is in the nontrivial phase. Based on the above discussion, we emphasize that unidirectional parts such as one-way fibers are necessary for the nonreciprocal generation of the Schr\"{o}dinger cat state in this scheme. They allow the CCW and CW modes in the $j$th cavity to couple with the CW mode of the ($j+1$th) cavity and CCW mode of the ($j-1$th) cavity, respectively. Consequently, the arrayed toroid cavity can be analogous to a typical topological SSH model, and the symmetrical CW and CCW modes in one single cavity can become different sublattices of the unit cell in the SSH structure. This is crucial for topological-protected chiral-mode excitation on the CW and CCW cavity modes in the edge cavity. Moreover, implementing such one-way fibers is quite difficult in a periodic structure because it involves many fibers. One should carefully adjust the direction and length of fibers as well as guarantee that the fiber and cavity are correctly contacted in every unit cell.
\begin{figure*}
	\centering
	\includegraphics[width=16cm]{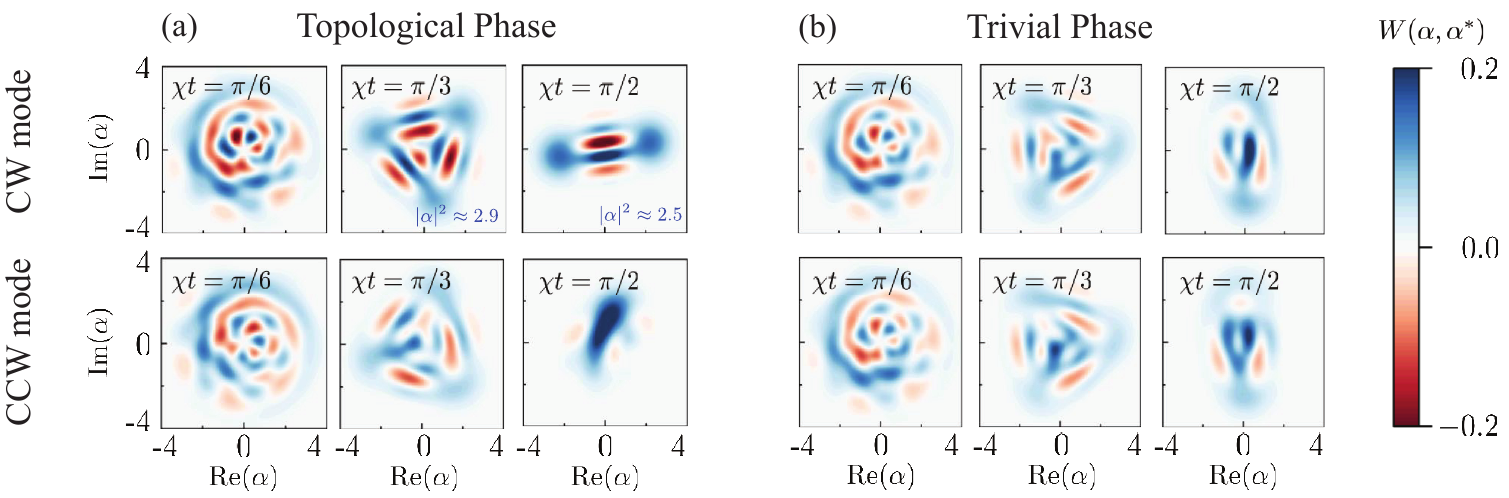}
	\caption{(Color online) Time evolution of the Wigner function of edge cavity modes in (a) topological phase ($ t_{1} = 0.4\kappa
		$, $ t_{2} = 8\kappa $) and (b) trivial phase ($ t_{1} = 6\kappa $, $ t_{2} = 4\kappa $). The upper row corresponds to the CW mode and the lower row corresponds to the CCW mode. The inserted $|\alpha|^2$ denotes the size of cat state. Other parameters are $\alpha_0=2$, $\chi = 2\pi\kappa $, and $ N=5 $.}\label{fig3}
\end{figure*}

Based on the above chiral mode excitation in the single edge cavity, the Kerr nonlinear effect will exhibit great chirality in the nontrivial phase, even if there is no difference between the Kerr strengths of the CW and CCW direction. To show this chiral Kerr effect, we drive the edge cavity from the opposite directions by a laser with power $ P_d $ and frequency $ \omega_d$, and investigate the steady-state excitation spectrum of this cavity [see Fig.\,\ref{fig2}(d)]. The driven Hamiltonian is $H_d=i\sqrt{2\gamma}(\epsilon e^{-i\omega_d t} a_{1,\circlearrowright/\circlearrowleft}^{\dagger}-h.c.)$, where $\epsilon=\sqrt{2\pi P_d/\hbar\omega_d}$ and $ \gamma $ is the loss induced by the extra driving fiber. Based on the master equation,
\begin{align}
	\dot{\rho} =& -i[H_{\rm tot}+H_d, \rho] + 2\gamma\mathcal{L}(a_{1,\circlearrowright})\rho+2\gamma\mathcal{L}(a_{1,\circlearrowleft})\rho \nonumber \\
	&+ \sum_{j}(2\kappa\mathcal{L}(a_{j,\circlearrowright})\rho+2\kappa\mathcal{L}(a_{j,\circlearrowleft})\rho).
\end{align}
we can obtain the steady-state density matrix $ \rho_s $. Here, $\rho$ is the density matrix of the total system, $\kappa$ is the total loss rate of the cavity mode without external driving, and $\mathcal{L}(a)\rho = a \rho a^{\dagger} - 1/2 \{a^{\dagger} a, \rho\} $. Then we can calculate the excitation spectrum of the driven edge cavity as $ \langle n_{\circlearrowright/\circlearrowright} \rangle = {\rm Tr}[ a^{\dagger}_{1,\circlearrowright/\circlearrowright}a_{1,\circlearrowright/\circlearrowright}\rho_s] $ with respect to different $ \Delta $, as shown in Fig.\,\ref{fig2}(c). In the topological phase, when the laser is driven from the CW direction, we can observed the obvious peak at the Kerr sideband $ \Delta=\omega_a-\omega_d=-\chi $ attributed to the two-photon resonance. This indicates the strong Kerr nonlinearity of the CW mode. The peak at the Kerr sideband turns into two separated peaks when the laser is driven from the CCW direction because the driving from the opposite direction excites bulk states, which have frequencies split away from the single-cavity resonant frequency. The disappearance of the Kerr sideband peak indicates that the Kerr nonlinearity of the CCW mode is strongly suppressed, then the chiral excitation at the Kerr sideband $\Delta=-\chi$ can serve as an observable quantity for featuring the chirality of the Kerr nonlinear effect. It is also shown in Fig.\,\ref{fig2}(d) that the chiral Kerr effect between the CW and CCW modes almost vanishes in the trivial phase. To view the detail variation of the chiral Kerr effect in the topological and trivial phases, we define the corresponding difference rate as $R_{\rm k}=|n_{\circlearrowright}-n_{\circlearrowleft}|/(n_{\circlearrowright}+n_{\circlearrowleft}) $, where $n_{\circlearrowleft/\circlearrowright}$ is the CW/CCW mode excitation number at $ \Delta=-\chi$, and plot $R_{\rm k}$ as a function of $t_1/t_2$ in Fig.\,\ref{fig2}(b). The variation of these rates in the topological and trivial phases further indicates that the topology can result in a great chiral Kerr effect in the nontrivial phase.

\emph{Nonreciprocal generation of the macroscopic quantum superposition state.}---The above nonreciprocal Kerr effect offers an opportunity for achieving the {\it nonreciprocal generation of the Schr\"{o}dinger cat state}. With the same initial condition, we can prepared a high-fidelity Schr\"{o}dinger cat state ({\it quantum state}) in the CW mode but only a classical state in the CCW mode. To investigate this nonreciprocal preparation, assume that the CW(CCW) mode on the left-edge cavity is initially in a coherent state and the other cavity field modes are initially in the vacuum states. To be specific, for the preparation on the CW direction, the initial state is $ | \Psi_\circlearrowright(0)\rangle = | \alpha_0 \rangle \otimes | 0 \rangle \otimes ... \otimes | 0 \rangle $, and for the preparation on the CCW direction, the initial state is $ | \Psi_\circlearrowleft(0)\rangle = | 0 \rangle \otimes | \alpha_0 \rangle \otimes ... \otimes | 0 \rangle $. This is a natural assumption for coherent input light injected into the cavity with a time scale shorter than $1/\kappa$. The time evolution of the system is based on the $ H_{\rm tot} $, which is described by the master equation
\begin{align}
	\dot{\rho} = -i[H_{\rm tot}, \rho] + \sum_{j}(2\kappa\mathcal{L}(a_{j,\circlearrowright})\rho+2\kappa\mathcal{L}(a_{j,\circlearrowleft})\rho).
\end{align}
In the topological nontrivial phase, the two states given by the time evolution will exhibit significant differences between the opposite directions. The numerical time evolutions of edge cavity modes in a 1D array with 5 cavities are displayed in the left panel of Fig.\,\ref{fig3}. For the state evolution of the CW mode, a three-component cat state with large size $ |\alpha|^2 = 2.9 $ at time $ \chi t = \pi/3 $ and a two-component cat state with large size $ |\alpha|^2 = 2.5 $ at time $ \chi t = \pi/2 $ are obtained. Recognizable macroscopical peaks and their quantum interference can be observed. These results can be explained by a single-mode time evolution with the Kerr effect because the edge state in the topological phase is almost completely localized on the CW mode of the edge cavity and the Kerr effect of the CW mode is effectively excited. In this single-mode time evolution, the initial state should be $ | \alpha_0 \rangle $, and the time evolution operator is given by
\begin{align}\label{eq6}
	U(t) = exp(-i(\omega_a a^{\dagger} a + \chi a^{\dagger}a^{\dagger} a a)t).
\end{align}
In the interaction picture($ \omega_a=0 $), the $ U(t) $ will give a state of the form $ | \psi \rangle = e^{-it [(a^{\dagger}a)^2-a^{\dagger}a] \chi} | \alpha_0 \rangle = e^{-|\alpha_0|^2/2}\sum_n{\alpha_0^n/\sqrt{n!}\;e^{-it (n^2-n) \chi}| n \rangle} $ at time $ t $. When $ \chi t = \pi/3 $, the state will evolve to the three-component cat state $|\psi \rangle = c_1 | \alpha_0 e^{-i(2\pi/3)} \rangle + c_2 | \alpha_0 \rangle + c_3 | \alpha_0 e^{i(2\pi/3)} \rangle $, with $ c_1 = (1-2e^{-i\pi/3})/3 $, $ c_2 = c_3 = (1 + e^{-i\pi/3})/3 $. At $ \chi t = \pi/2 $, the state becomes $ | \psi \rangle = 1/\sqrt{2}(e^{i\pi/4} | -i\alpha_0 \rangle+e^{-i\pi/4} | i\alpha_0 \rangle) $, which is a two-component cat state. However, for the CCW case, the state evolutions are apparently different, and only the states with low nonclassicality are obtained at the same time. This can be explained by the suppression of Kerr effect in the CCW mode owing to chiral-mode excitation. Finally, the obtained state can be extracted through a quantum state transferring protocol\,\cite{Xiang2017}. We note that if we inject the input light and extract the final state by the same extra fiber coupled to the cavity, then the input field in one direction will only result in the extracted state in the opposite direction. And the same coherent input fields in the two directions will result in different extracted states (a quantum state and a classical state) after the same period of time. Thus our proposed model can give the nonreciprocal generation of the cat state in the topological phase. Furthermore, this quantum nonreciprocity reduces significantly in the trivial phase. As shown in the right panel of Fig.\,\ref{fig3}, the generations of cat states perform insignificant distinction for the CW and CCW modes at the same time. For our chosen parameters, neither the CW nor CCW mode can evolve into a high-quality cat state. This result can be understood as follows: the nonreciprocal Kerr effect is substantially destructed when the microcavity array enters into the trivial phase, and thus, the above quantum nonreciprocity vanishes.

\begin{figure}
	\centering
	\includegraphics[width=8.5cm]{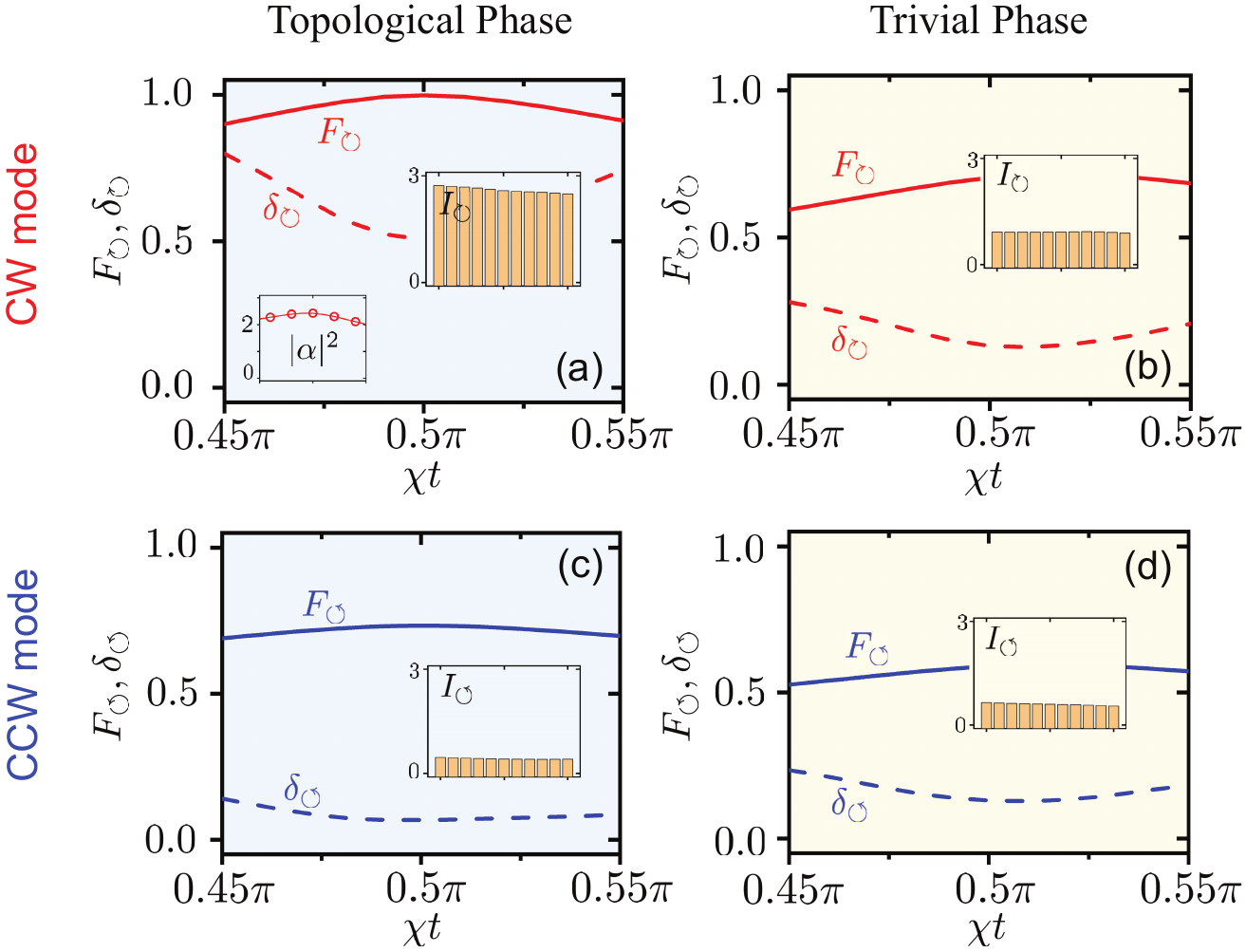}
	\caption{(Color online) Time evolution of fidelity $F$, Wigner negativity $ \delta $, and macroscopic quantum superposition $I$ (inset) around $ \chi t=\pi/2 $ in the topological phase (a,c), i.e., $ t_{1} = 0.4\kappa $, $ t_{2} = 8\kappa $, and the trivial phase (b,d), i.e., $ t_{1} = 6\kappa $, $ t_{2} = 4\kappa $. The inserted $|\alpha|^2$ in (a) indicates the size of obtained cat state, and other parameters are the same as those in Fig.\,\ref{fig3}.
	}\label{fig4}
\end{figure}

In order to quantify the nonreciprocity of generation more clearly, we calculate the figures of merit for the two-component cat states, and compare their differences between the CW and CCW modes in Fig.\,\ref{fig4}. Here $\delta$ is the Wigner negativity to evaluate the nonclassicality of the state. $I$ is used to quantify the macroscopic quantum superposition, and a perfect cat state has the maximum value $ I = \langle a^{\dagger}a \rangle $. $F$ is the fidelity between the actual state and the most similar cat state. To be specific, the most similar cat state takes the form $ | \psi_{\sigma}(\eta,\beta,\gamma) \rangle = \mathcal{N}(| -\eta e^{i\gamma} \rangle+e^{i\beta}| \eta e^{i\gamma} \rangle) $, where $\mathcal{N}$ is the normalization factor, $ \eta $, $ \beta $, and $ \gamma $ are the fitting coefficients. We can obtain $ \eta $, $ \beta $, and $ \gamma $ from the optimization calculation described by
\begin{align}
	&\mathop{\rm min}_{\eta,\beta,\gamma}(\int [ W_{|\psi_{\rho} \rangle}(\bm{\alpha}) - W_{|\psi_{\sigma}(\eta,\beta,\gamma) \rangle}(\bm{\alpha}) ]^2d^{2}\bm{\alpha}),\nonumber \\
	&s.t. \quad \alpha \in [ 1.5, +\infty ), \beta \in [-\pi, +\pi ], \gamma \in [-\pi, +\pi ],
\end{align}
where  $ |\psi_{\rho} \rangle $ is the actual obtained state, and $W_{|\psi \rangle}(\bm{\alpha})$ is the Wigner function of state $ |\psi \rangle $. Physically, the optimized state $ | \psi_{\sigma}(\eta,\beta,\gamma) \rangle$ has the most similar profile to the actual state. The fidelity of the actually obtained state is measured with respect to the most similar cat state instead of the theoretical state $|\psi \rangle$ given by Eq.~(\ref{eq6}), because the dissipation in the realistic environment will greatly modify the phase and size of the actually obtained state. In some cases, the actually obtained state has little overlap with $|\psi \rangle$, but still can be treated as the superposition of two macroscopic coherent states. The Wigner negativity, macroscopic quantum superposition, and fidelity are defined as
\begin{align}
	&\delta = \int (| W(\bm{\alpha}) |-W(\bm{\alpha}))d^{2}\bm{\alpha},
\end{align}
\begin{align}
	&I = \frac{\pi}{2}\int W(\bm{\alpha})(-\frac{\partial^{2}}{\partial\alpha\partial\alpha^{*}}-1)W(\bm{\alpha})d^{2}\bm{\alpha},
\end{align}
\begin{align}
	&F(\rho,\sigma) = {\rm Tr}\sqrt{\sqrt{\rho}\sigma\sqrt{\rho}}\sim|\langle \psi_{\rho}|\psi_{\sigma} \rangle|,
\end{align}
where $ \rho $ and $ \sigma $ are the density matrix of the obtained state $ |\psi_{\rho} \rangle $ and the most similar cat state $ |\psi_{\sigma} \rangle $, respectively. Firstly, it is shown from Fig.\,\ref{fig4}(a) that, in the topological phase, the obtained two-component cat state for CW mode has high nonclassicality ($\delta_{\circlearrowright} \approx 0.5$), fidelity ($F_{\circlearrowright}\to1$), macroscopic quantum superpositions ($I_{\circlearrowright} \approx 2.6$), and large size ($|\alpha|^2 \approx2.5$), which are enough for practical applications in quantum technologies\,\cite{Joo2011, Ralph2003, Mirrahimi2014}. Secondly, the above key properties of the cat state also imply the strong nonreciprocity of the cat state generation in the topological nontrivial phase. Comparing Fig.\,\ref{fig4}(a) and Fig.\,\ref{fig4}(c), one can find that the nonclassicality is exhibited in the CW mode and vanishes in the opposite direction. The Wigner negativities (dashed line) of the CW and CCW modes are $ \delta_{\circlearrowright} \approx 0.5$ and $ \delta_{\circlearrowleft} \approx 0.06 $ at time $ \chi t = \pi/2 $, respectively. The fidelities (solid line) of the CW and CCW modes are $F_{\circlearrowright} \to 1$ and $ F_{\circlearrowleft} \approx 0.73 $ at time $ \chi t = \pi/2 $, respectively, which indicates the great difference of the quality of cat states obtained in the two directions. The different quantum coherence signified by the macroscopic quantum superpositions $I$ is shown in the inset with $ I_{\circlearrowright} \approx 6 I_{\circlearrowleft}$. A more clear nonreciprocal rate can be defined based on fidelity as $ R_{\rm F}=|F_{\circlearrowright}-F_{\circlearrowleft}| $, and $ R_{\rm F}=0.27 $ is found for the topological case. Lastly, it is shown from Fig.\,\ref{fig4}(b,d) that the nonreciprocity implied by the above key properties of the cat state is significantly reduced in the trivial phase. The Wigner negativity, macroscopic quantum superpositions, and fidelity indicate that both the CW and CCW modes cannot evolve into high-quality cat states in the trivial phase. Especially, the fidelity for CW mode is $ F_{\circlearrowright} \approx 0.7 $ and for CCW mode is $ F_{\circlearrowleft} \approx 0.59 $, the nonreciprocal rate is then reduced to $ R_{\rm F}=0.11 $. This is attributed to the disappearance of chiral mode excitation in the trivial phase. In the supplementary material, we show that the nonreciprocal generation of a three-component cat state has similar results.

\emph{Discussion and Conclusion.}---For experimental implementations, the resonator array with an ultrahigh Q ($ Q \sim 10^8 $) microtoroid cavity has been demonstrated in Refs.\,\cite{Armani2003,Ostby2009}. The required unidirectional fibers that connect cavities are achievable through chiral interactions\,\cite{Lodahl2017} or spatiotemporal modulation of material\,\cite{Yu2009}. Based on the development of microcavity photonics, the strong Kerr nonlinearity for the microresonator $ \chi\approx5$MHz has been achieved experimentally\,\cite{Palomo2017,Zielinska2017}. Tunable backscattering interaction that can reach to $ t_1\approx3$MHz was experimentally demonstrated by modifying the distance between the scatterer and resonator\,\cite{Zhu2010,Peng2016}. Meanwhile, in the $\rm Er^{3+} $-doped microcavity, a gain with rate $\mathcal{G}$ can be introduced by optically pumping $ \rm Er^{3+} $ ions with a pump laser, and thus the effective loss rate $\kappa$ of the cavity mode becomes tunable\,\cite{Peng2014,Chang2014}. In this case, the required coupling-loss relation $t_{1,2}/\kappa$ and nonlinearity-loss relation $\chi/\kappa$ can be achieved by employing the proven $\rm Er^{3+}-$doping technologies in our proposal.

In summary, we have proposed an experimentally feasible scheme to realize the {\it nonreciprocal generation of the Schr\"{o}dinger cat state}, and revealed the relation between this quantum nonreciprocity and topology in a 1D microtoroid cavity array. This {\it nonreciprocal generation of macroscopic quantum superposition} emerges in the edge cavity when the microcavity array enters into the topological nontrivial phase. We have also calculated the key properties of cat state, including the Wigner negativity, macroscopic quantum superposition and fidelity, and all of them imply that the generations of cat state exhibit unidirectional high qualities in the topological nontrivial phase. This guarantees that the obtained nonreciprocal generation of the cat state, as a pivotal quantum resource, can be used to implement one-way quantum information processing with high efficiency. Our work opens up a door for exploring a new type of nonreciprocal quantum resources, together with their applications in quantum information science and photonics by utilizing the topology of system.

\let\oldaddcontentsline\addcontentsline% Store \addcontentsline
\renewcommand{\addcontentsline}[3]{}% Make \addcontentsline a no-op

\let\addcontentsline\oldaddcontentsline% Restore \addcontentsline

%%%%%%%%%% Merge with supplemental materials %%%%%%%%%%
\onecolumngrid

%%%%%%%%%% Prefix a "S" to all equations, figures, tables and reset the counter %%%%%%%%%%
\newcommand\specialsectioning{\setcounter{secnumdepth}{-2}}
\setcounter{equation}{0} \setcounter{figure}{0}

\setcounter{table}{0}
\renewcommand{\theequation}{S\arabic{equation}}
\renewcommand{\thefigure}{S\arabic{figure}}
\renewcommand{\bibnumfmt}[1]{[S#1]}
\renewcommand{\citenumfont}[1]{S#1}
\renewcommand\thesection{S\arabic{section}}
%%%%%%%%%% Prefix a "S" to all equations, figures, tables and reset the counter %%%%%%%%%%
\renewcommand{\baselinestretch}{1.2}

%\renewcommand{\theequation}{S\arabic{equation}}

%%%%%%%%%%%%%%%%%%%%%%%%%%%%%%%%%%%%%%%%%%%%%%%%%%%%%%%%%%%%%%%%%
\newpage

\setcounter{page}{1}\setcounter{secnumdepth}{3} \makeatletter
\begin{center}
{\Large \textbf{ Supplemental Material for\\
        “Nonreciprocal Generation of Schr\"{o}dinger Cat State Induced by Topology”}}
\end{center}

\begin{center}
Zi-Hao Li$^{1}$, Li-Li Zheng$^{2}$, Ying Wu$^{1}$, Xin-You L\"{u}$^{1}$
\end{center}

\begin{minipage}[]{16cm}
\small{\it
	\centering $^{1}$School of Physics and Institute for Quantum Science and Engineering, Huazhong University of Science and Technology, Wuhan 430074, China \\}
\small{\it
	\centering $^{2}$Key Laboratory of Optoelectronic Chemical Materials and Devices of Ministry of Education, Jianghan University, Wuhan 430074, China \\}
\end{minipage}

\vspace{8mm}

This supplemental material contains six parts:  I. A detailed derivation for the effective link Hamiltonian $\tilde{H}_l$ used in the main text. II. An additional analysis for nonreciprocal generation of three-component cat State. III. Discussion of nonreciprocal performance for different number of cavities. IV. An additional discussion for the influence of cross-Kerr nonlinearity. V. Some theoretical analyses for the nonreciprocal transmission of optical field. VI. Experimental feasibility of our proposal. 

%%%%%%%%%%%%%%%%%%%%%%%%%%%%%%%%%%%%%%%%%%%%%%

\section{Derivation of effective link Hamiltonian}\label{sectionI}

\begin{figure}
	\includegraphics[width=14cm]{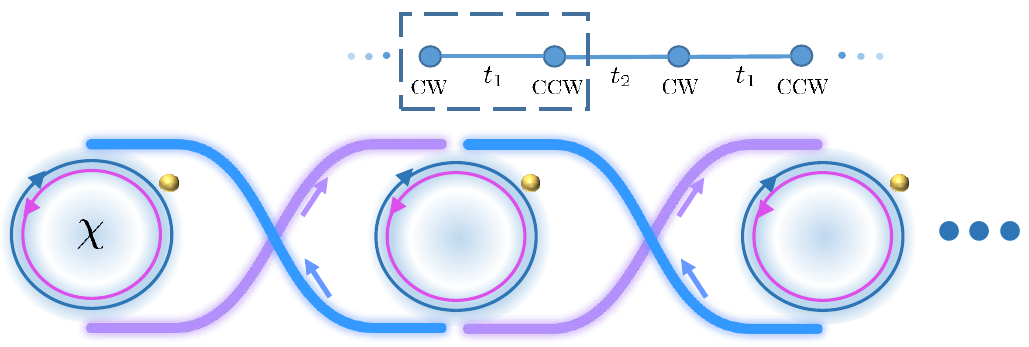}
	\caption{A 1D micro-toroid cavity array where each cavity supports two counter-circulating modes that are coupled with each other via a scatterer. The adjacent cavities are connected by the unidirectional fibers. The cavity array can be described by a SSH model shown in the right top. The dashed box is a unit cell that consists of the CW and CCW modes in one cavity. Here $t_{1}$, $t_{2}$ are the on-site and off-site coupling strengths, respectively. The nonreciprocal generation of Schr\"{o}dinger cat states are achieved in the left edge cavity with a Kerr nonlinearity $\chi$.}
	\label{fig1_sm}
\end{figure}

In this section, we briefly describe the derivation of effective link Hamiltonian $\tilde{H}_l$ by eliminating the fiber modes. We consider that the $ j $th cavity and the $ (j+1) $th cavity are connected by two unidirectional fibers, as shown in Figs.\,\ref{fig1_sm}, and the link Hamiltonian reads
\begin{align}
	\!\!H_{l} =& i\sqrt{2\kappa_f}[a_{j,\circlearrowleft}f^{\dagger}_{j,\rm R}(t,z_j)+a_{j,\circlearrowleft}\mathscr{F}^{\dagger}_{j,\rm L}(t,z_j)+a_{j+1,\circlearrowright}f^{\dagger}_{j,\rm R}(t,z_{j+1})+a_{j+1,\circlearrowright}\mathscr{F}^{\dagger}_{j,\rm L}(t,z_{j+1})\!-\!h.c.],
\end{align}
where $ \kappa_f $ is the loss induced by the resonator-fiber coupling, and $ z_{j} $ is the cavity position. Here, $f_{j, \rm R}(t,z) $ is the right propagating fiber mode in $j$th purple fiber. $\mathscr{F}_{j, \rm L}(t,z) $ is the left propagating fiber mode in $j$th blue fiber, where the $j$th fiber is placed on the right of $j$th cavity. They are represented by a continuum of left and right propagating bosonic mode, i.e., $ f_{j,R}(t,z) = \frac{1}{\sqrt{2\pi}}\int_0^\infty d\omega f_{j,\omega} e^{-i\omega (t - z/c)} $, $ \mathscr{F}_{j,\rm L}(t,z) = \frac{1}{\sqrt{2\pi}}\int_0^\infty d\omega \mathscr{F}_{j,\omega} e^{- i\omega (t + z/c)} $ and $ [f_{j,\omega}, f^{\dagger}_{j,\omega^{\prime}}]=\delta(\omega-\omega^{\prime}) $, $ [\mathscr{F}_{j,\omega}, \mathscr{F}^{\dagger}_{j,\omega^{\prime}}]=\delta(\omega-\omega^{\prime}) $. In principle, it can be considered as two cascading systems\,\cite{Gardiner2004, Bin2018, Cirac1997, Xiang2017, Stannigel2010, Stannigel2011}. We first focus on the cascading system that describes the CCW mode in $j$th cavity and CW mode in $(j+1)$th cavity connected by the right propagating fiber, and its Hamiltonian is
\begin{eqnarray}
	H_{\rm R} = i\sqrt{2\kappa_f}(a_{j,\circlearrowleft}f^{\dagger}_{j,\rm R}(t,z_j)+a_{j+1,\circlearrowright}f^{\dagger}_{j,\rm R}(t,z_{j+1})-h.c.).
\end{eqnarray}
We can eliminate the fiber mode under the Born-Markov approximation, and define the input and output field operators in the form $f^{\rm in}_{j,R}(t,z_j) = f_R(t,z_j+0^-)$ and $f^{\rm out}_{j,R}(t,z_j) = f_{j,R}(t,z_j+0^+) $, respectively. This gives the dissipative dynamics in terms of quantum Langevin equation (QLE) of CCW mode in $j$th and CW mode in ($j+1$)th cavity
\begin{eqnarray}
	\dot{a}_{j,\circlearrowleft} &=& - \kappa_f a_{j,\circlearrowleft} - \sqrt{2\kappa_f} f_{j,\rm R}^{\rm in}(t,z_{j}),\nonumber\\ 
	\dot{a}_{j+1,\circlearrowright} &=& - \kappa_f a_{j+1,\circlearrowright} - \sqrt{2\kappa_f} f_{j,\rm R}^{\rm in}(t,z_{j+1}), 
\end{eqnarray}
together with the input-output relation
\begin{eqnarray}
	f_{j,\rm R}^{\rm out}(t,z_{j}) &=& f_{j,\rm R}^{\rm in}(t,z_{j}) + \sqrt{2\kappa_f} a_{j,\circlearrowleft}(t),\nonumber\\ 
	f_{j,\rm R}^{\rm out}(t,z_{j+1}) &=& f_{j,\rm R}^{\rm in}(t,z_{j+1}) + \sqrt{2\kappa_f} a_{j+1,\circlearrowright}(t). 
\end{eqnarray}
In this cascading system, $ f_{j,\rm R}^{\rm in}(t,z_{j}) $ is a $ \delta $-correlated noise operator representing the white quantum noise. From the unidirectional feature of the fiber, the output of $j$th CCW mode can travel through fiber and become the input of the $(j+1)$th CW mode with a time delay, which is described by the following form\,\cite{Stannigel2010, Stannigel2011}
\begin{eqnarray} \label{S7}
	f_{j,\rm R}^{\rm in}(t,z_{j+1}) = f_{j,\rm R}^{out}(t-\tau,z_{j})e^{i\omega \tau}.
\end{eqnarray}
Here, $ \tau=(z_j-z_{j+1})/c $ is the propagation time between the neighboring cavities. To obtain a simple form, we define the retard operators $ a_j(t)\rightarrow a_{j}(t-z_j/c) $. The term $ e^{i\omega \tau} $ can be settled to imaginary unit $ i $ by tuning the positions of cavity along the fiber to satisfy the relation $ z_j-z_{j+1} = \pi c/(2\omega)$. As a result, Eq.\,(\ref{S7}) becomes
\begin{eqnarray}
	f_{j,\rm R}^{\rm in}(t,z_{j+1}) = i f_{j,\rm R}^{\rm out}(t,z_j).
\end{eqnarray}

For the cascading system that describes the CW mode in $ j $th cavity and CCW mode in $(j+1)$th cavity connected by the left propagating fiber, the relevant Hamiltonian is
\begin{eqnarray}
	H_{\rm L} = i\sqrt{2\kappa_f}(a_{j,\circlearrowleft}\mathscr{F}^{\dagger}_{j,\rm L}(t,z_j)+a_{j+1,\circlearrowright}\mathscr{F}^{\dagger}_{j,\rm L}(t,z_{j+1})-h.c.).
\end{eqnarray}
We can obtain the similar QLEs and input-output relations. We also consider the input fields are quantum white noises written in the form
\begin{eqnarray}
	f^{\rm in}_{j,R}(t,z_{j}) = \mathscr{F}^{\rm in}_{j,L}(t,z_{j+1}) = dB(t),
\end{eqnarray}
with
\begin{eqnarray}
	&&[dB(t)]^2=[dB^{\dagger}]^2=0,\nonumber\\ 
	&&dB(t)dB(t)^{\dagger}=(N+1)dt,\\
	&&[dB]^{\dagger}dB(t) = Ndt \nonumber
\end{eqnarray}

From this we can derive a master equation for density operator $ \rho(t) $ of the two cavity system by setting $\langle da(t)\rho \rangle \equiv \langle ad\rho(t) \rangle $, and the master equation can be written in the form
\begin{eqnarray}\label{S9}
	\dot{\rho} =& 2\kappa_f [\mathcal{L}(a_{j,\circlearrowright/\circlearrowleft})\rho+\mathcal{L}(a_{j+1,\circlearrowright/\circlearrowleft})\rho]+2\kappa_f\mathcal{L}_{\rm noise}(a_{j,\circlearrowright/\circlearrowleft},a_{j+1,\circlearrowright/\circlearrowleft})\rho-2i\kappa_f([a_{j+1,\circlearrowright}^{\dagger},a_{j,\circlearrowleft}\rho]-[\rho a_{j,\rm \circlearrowleft}^{\dagger},a_{j+1,\rm \circlearrowright}]).
\end{eqnarray}

Here, $ \mathcal{L}(a)\rho = a \rho a^{\dagger} - 1/2 \{a^{\dagger} a, \rho\} $ is the generalized Lindblad superoperator and $ \mathcal{L}_{\rm noise}(a,b)\rho = N/2 ([[a+b,\rho],a^{\dagger}+b^{\dagger}]+h.c.) $. The first term of Eq.\,(\ref{S9}) describes the decay of cavity introduced by cavity-fiber coupling. In the case that $ N=0 $, which means the input fields are vacuum fields, the second term can be safely eliminated. Then we can obtain a stochastic Schr\"{o}dinger equation for the cascaded systems from the master equation. The evolution of stochastic wave function is written as
\begin{eqnarray}
	\frac{d}{dt} | \Psi(t) \rangle = -i H_{l}^{\rm eff} | \Psi(t) \rangle,
\end{eqnarray}
where
\begin{eqnarray}
	H_{l}^{\rm eff} = (t_2 a_{j,\rm \circlearrowleft}^{\dagger}a_{j+1,\rm \circlearrowright}+t_2 a_{j+1,\rm \circlearrowright}^{\dagger}a_{j,\rm \circlearrowleft})
\end{eqnarray}
represents the effective cascading coupling between $ j $th and $(j+1)$th cavity and $ t=2\kappa_f $ is the coupling coefficient. By counting all the cascading couplings in the 1D cavity array, the effective link Hamiltonian of the total system reads
\begin{eqnarray}
	\tilde{H}_{l} = \sum^{N-1}_{j-1}  (t_2 a_{j,\rm \circlearrowleft}^{\dagger}a_{j+1,\rm \circlearrowright}+t_2 a_{j+1,\rm \circlearrowright}^{\dagger}a_{j,\rm \circlearrowleft}).
\end{eqnarray}

\section{Additional analysis for the nonreciprocal generation of three-component cat state} 
In the main text, it is shown that the nonreciprocal generation of three-component Schr\"{o}dinger cat states with large size can be achieved at $\chi t = \pi/3$. Here we provide some detail analysis to quantify the characters together with the nonreciprocity of the cat states generation by applying three figures of merit, i.e., Wigner negativity $ \delta $, macroscopic quantum superpositions $ I $ and fidelity $ F $, defined as
\begin{align}
	&\delta = \int (| W(\bm{\alpha}) |-W(\bm{\alpha}))d^{2}\bm{\alpha},
\end{align}
\begin{align}
	&I = \frac{\pi}{2}\int W(\bm{\alpha})(-\frac{\partial^{2}}{\partial\alpha\partial\alpha^{*}}-1)W(\bm{\alpha})d^{2}\bm{\alpha},
\end{align}
\begin{align}
	&F(\rho,\sigma) = {\rm Tr}\sqrt{\sqrt{\rho}\sigma\sqrt{\rho}}\sim|\langle \psi_{\rho}|\psi_{\sigma} \rangle|.
\end{align}
\begin{figure}[t]
	\includegraphics[width=12cm]{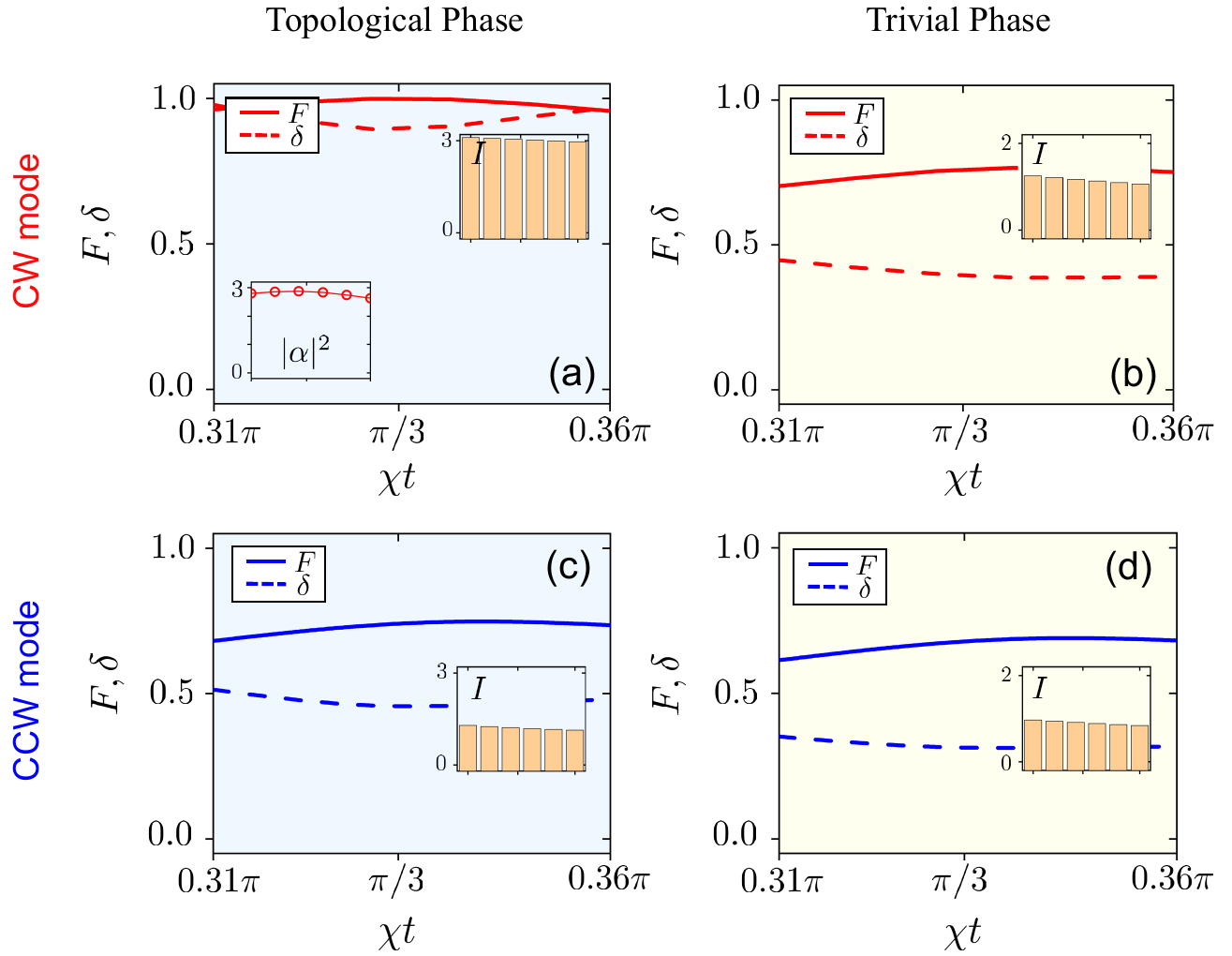}
	\caption{Time evolution of Fidelity $F$, Wigner negativity $ \delta $ and macroscopic quantum superposition $I$ (inset) around $ \chi t=\pi/3 $ in the topological phase (a, c), i.e., $ t_{1} = 0.4\kappa $, $ t_{2} = 8\kappa $, and the trivial phase (b, d), i.e., $ t_{1} = 6\kappa $, $ t_{2} = 4\kappa $. The inserted $|\alpha|^2$ in (a) indicates the size of obtained cat state, and the other parameters are $\alpha_0=2$, $\chi = 2\pi\kappa $, $ N=5 $.}
	\label{fig2_sm}
\end{figure}
Here $W(\bm{\alpha})$ is the Wigner function of state, $ \rho $ and $ \sigma $ are the density matrix of obtained state $ |\psi_{\rho} \rangle $ and the most similar cat state $ |\psi_{\sigma} \rangle $, respectively. Fig\,\ref{fig2_sm}(a) show that, in topological nontrivial phase, the obtained three-component cat state with size $ |\alpha|^2 = 2.9 $ in the CW mode at $ \chi t = \pi/3 $ approaches high nonclassicality ($\delta_{\rm CW} \approx 0.9$), fidelity ($F_{\rm CW}\to1$) and macroscopic quantum superpositions ($I \approx 3$) that are enough for practical applications. As predicted by the chirality mode occupation in the topological phase, strong nonreciprocity appears when we compare two counterpropagating modes, see Fig.\,\ref{fig2_sm}(a) and \ref{fig2_sm}(c). At time $ \chi t = \pi/3 $, the difference of nonclassicality is revealed by the Wigner negativity (dashed line) that reduces from $ \delta_{\rm CW} \approx 0.9 $ in CW mode to $ \delta_{\rm CCW} \approx 0.5 $ in CCW mode. The quality of cat state in CW mode is also obviously superior to that in CCW mode, as indicated by the Fidelity (solid line) of CW $F_{\rm CW} \to 1$ and CCW modes $ F_{\rm CCW} \approx 0.75$. The macroscopic quantum superposition $I$ presented in the inset reaches $ I_{\rm CW} \approx 2 I_{\rm CCW}$, showing the different quantum coherence of counterpropagating modes. Compare to the two-component cat state, the nonreciprocity of three-component cat state generation is slightly reduced because of the short evolution time. In the trivial phase shown by Figs.\,\ref{fig2_sm}(b) and \ref{fig2_sm}(d), the nonreciprocity almost disappears as displayed by the insignificant difference of figures of merit between two counterpropagating modes. This is due to the disappearance of chirality mode occupation.

\section{Comparation between different number of cavities}
\begin{figure}[t]
	\centering
	\includegraphics[width=14cm]{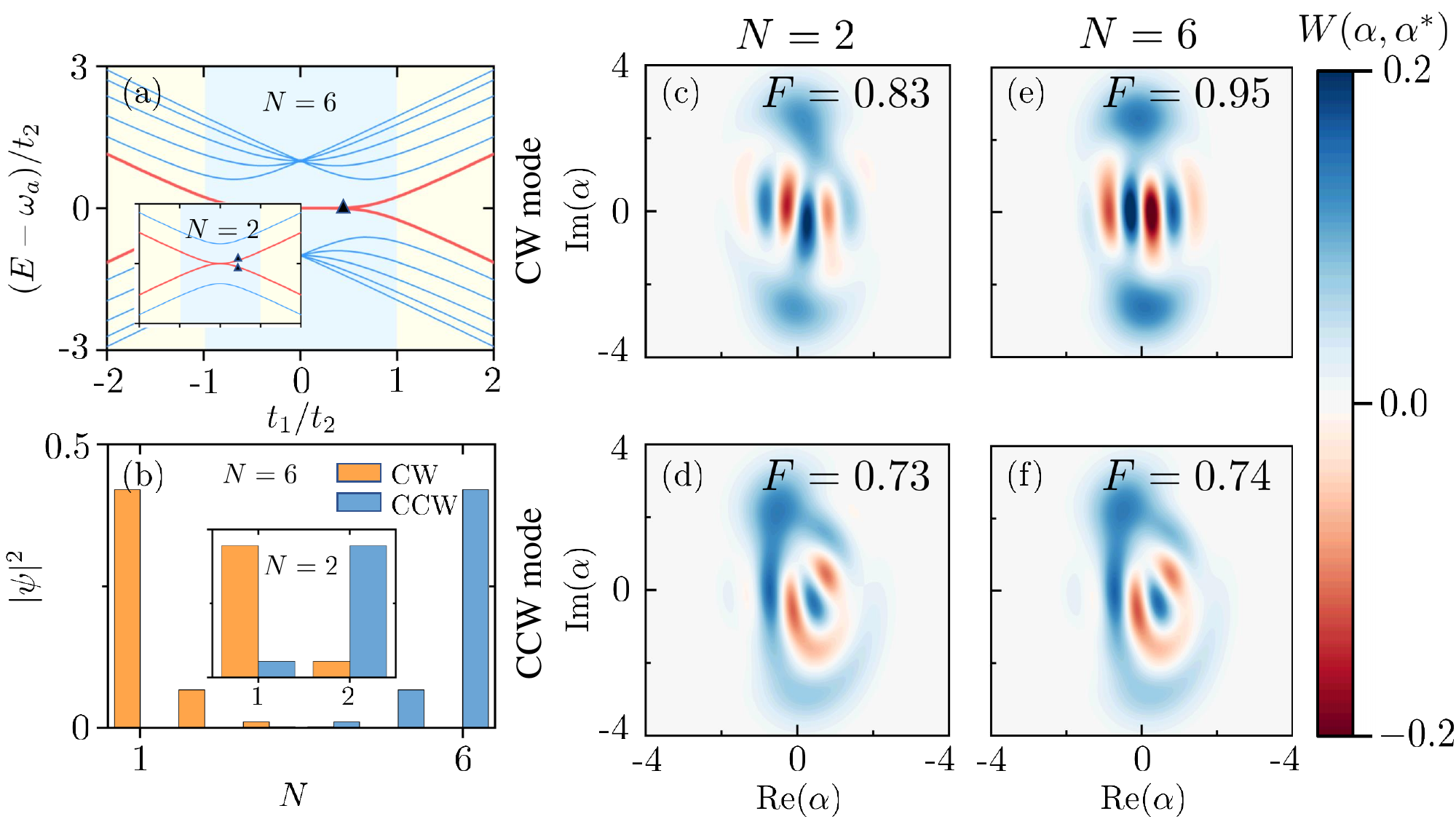}
	\caption{(Color online) (a) Energy spectrals of an open array with 6 cavities and 2 cavities (inset) versus $t_{1}/t_{2}$, The blue and yellow background indicate the topological nontrivial and trivial region given by the winding number of system. The red lines mark the edge modes. The black triangles indicate that the edge modes at $ t_1/t_2=0.4 $ degenerate for 6 cavities but split for 2 cavities. (b) The profiles of edge states at $ t_1/t_2 = 0.4 $ of an array with 6 cavities and 2 cavities (inset). (c)(d) The Wigner function of the edge cavity states obtained at $\chi t = \pi/2$ after the time evolution in an array with 2 cavities. (e)(f) The Wigner function obtained in the same condition, but in an array with 6 cavities. (c)(e) corresponds to the CW mode and the (d)(f) corresponds to the CCW mode. Other parameters are $t_1=8\kappa$, $t_2 = 20\kappa $, $\alpha_0=2$, $\chi = 4\pi\kappa $.}\label{fig3_sm}
\end{figure}
In this section we compare the nonreciprocal preparation of two-component cat state between different number of cavities. We first consider the cavity array for zero Kerr nonlinearity. The energy spectras of an array with 6 cavities and 2 cavities are plotted in Fig.\,\ref{fig3_sm}(a). Here, the closed energy gap corresponds to the appearance of edge mode. We can see that for both cases, the energy gap close at $|t_1/t_2|<1$, and the closing point for 2 cavities is closer to zero point compare to the closing point for 6 cavities. The closing point does not match the phase transition point given by the winding number because the size of system is limited. The topology is a many-body physical property of the system, thus in realistic system the topological effect will diminish when the size of system is decreased. As we can see at $ t_1/t_2=0.4 $, the closed energy gap for 6 cavities will open for 2 cavities. In Fig.\,\ref{fig3_sm}(b), we plot the profiles of ground states at $t_1/t_2=0.4$ in arrays with different sizes. In both arrays, we can see that the CW mode dominate the left edge and the CCW mode dominate the right edge. However, the chiral-mode-excitation at the edge cavity in an array with 6 cavities is much greater than that in an array with 2 cavities. This is a clear signal that the chiral-mode-excitation in our model is topologically protected, and the perfect chiral-mode-excitation is expected in a sufficiently large array.

To compare the nonreciprocities of cat state generations at $t_1/t_2=0.4$ between arrays of different cavites, we simulate the numerical time evolutions of the array with 2 cavities and 6 cavities following the master equation Eq.\,(5) in the main text, and the Wigner functions of state obtained at time $\chi t=\pi/2 $ are displayed in Fig.\,\ref{fig3_sm}(c)-(f). Here, the parameters are chosen as $t_1=8\kappa$, $t_2 = 20\kappa $ and $\chi = 4\pi\kappa $. Based on the discussion in the main text, the system is expected to give the nonreciprocal generation of cat state at this time. For arrays with different sizes, the fidelies of states in CCW mode are $ F=0.73 $ for 2 cavities and $F=0.74$ for 6 cavities, which means the qualities of states are low and almost indistinguishable for both cases. The difference can be found in CW mode, where the fidelity of states for 6 cavities is $ F=0.95 $ and for 2 cavities the fidelity is $ F=0.83 $. The quality for 6 cavities is obvioudly better than the quality for 2 cavities. As a result, the nonreciprocity of cat state generation for 6 cavities characterized by the nonreciprocal rate $ R_F(N=6)=0.12 $ is higher than the nonreciprocity for 2 cavities characterized by $ R_F(N=2)=0.06 $. The behavior of growing nonrecirpocity as the increase of system size is clear evidence that this nonreciprocity is a consequence of the topological effect.

\section{Influence of cross-kerr nonlinearity}
\begin{figure}[t]
	\centering
	\includegraphics[width=14cm]{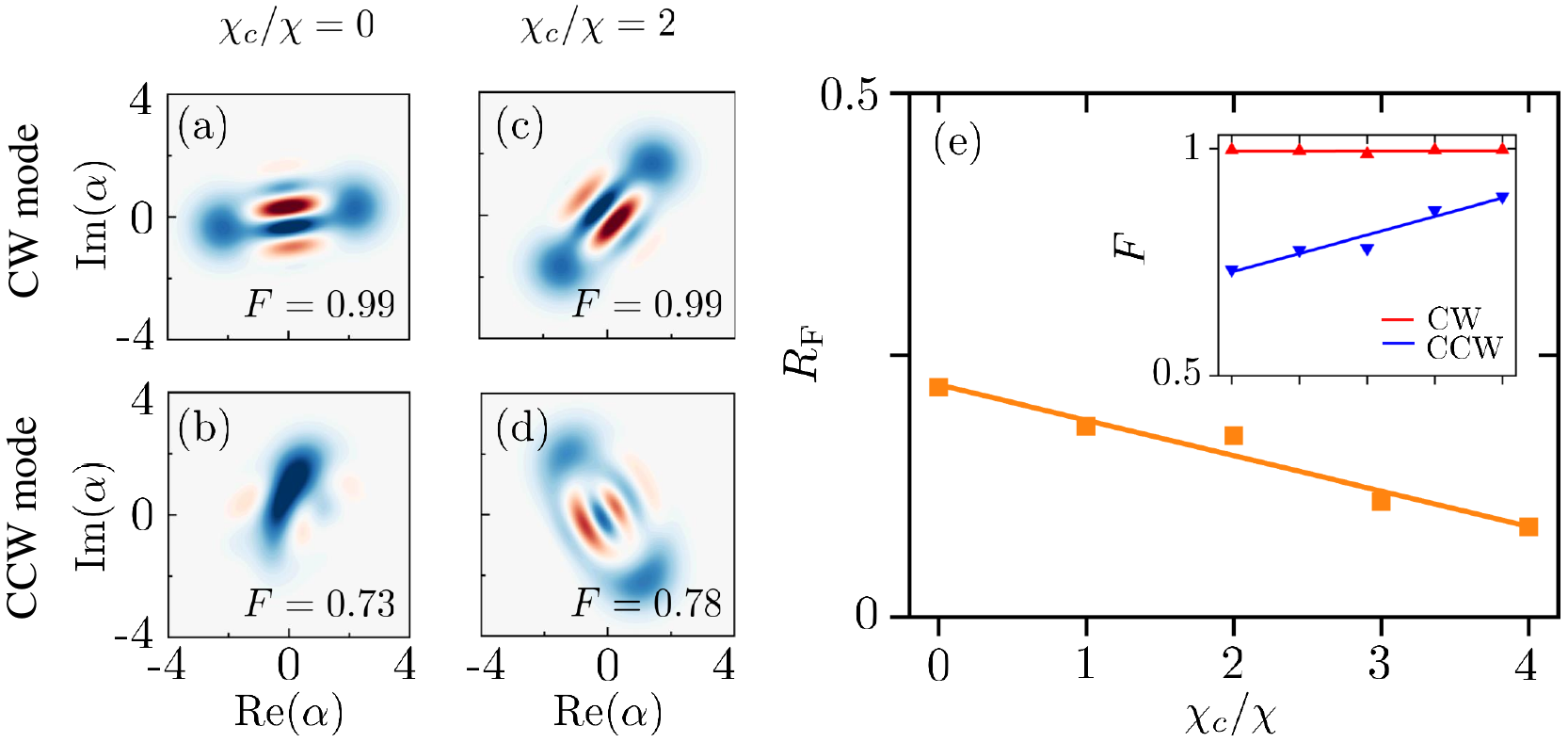}
	\caption{(Color online) (a)(b) Wigner functions of the obtained edge cavity states with cross-Kerr nonlinearity $ \chi_c/\chi = 0 $. (c)(d) The Wigner functions of the obtained edge cavity states with cross-Kerr nonlinearity $ \chi_c/\chi = 2 $. (a)(c) corresponds to the CW mode and the (b)(d) corresponds to the CCW mode. (e) The nonreciprocal rates based on Fidelity of obtained states $R_{\rm F} = |F_{\circlearrowright} - F_{\circlearrowleft}|$ with different cross-Kerr nonlinearity. The inset show the Fidelities of obtained state in CW and CCW modes. The lines represent the curve fitting of dots. Other parameters are $t_1=0.4\kappa$, $t_2 = 8\kappa $, $\alpha_0=2$, $\chi = 2\pi\kappa $, $\chi t = \pi/2$.}\label{fig4_sm}
\end{figure}
Generally, the Kerr nonlinearity in the edge toroid cavity can be introduced by various methods. However, some methods such as using optomechanical coupling will cause the cross-Kerr nonlinearity between counter-propagating modes\,\cite{Gong2009, Qian2021}. In this section, we discuss the influence of the cross-Kerr nonlinearity on the nonreciprocal generation of quantum states. We introduce the cross-Kerr Hamiltonian on the edge cavity of the form
\begin{align}
	H_{\rm CK} = \chi_{\rm c} a_{1,\circlearrowright}^{\dagger}a_{1,\circlearrowleft}^{\dagger}a_{1,\circlearrowright}a_{1,\circlearrowleft},
\end{align}
where $\chi_{\rm c}$ is the cross-Kerr nonlinear coefficient. In Fig.\,\ref{fig4_sm}, we numerically simulate the time evolutions following the master equation
\begin{align}
	\dot{\rho} = -i[H_{\rm tot}+H_{\rm CK}, \rho] + \sum_{j}(2\kappa\mathcal{L}(a_{j,\circlearrowright})\rho+2\kappa\mathcal{L}(a_{j,\circlearrowleft})\rho),
\end{align}
and calculate the fidelities of obtained states for different cross-Kerr nonlinear strength after the nonreciprocal generation processes. In Fig.\,\ref{fig4_sm}(a)(b), the Wigner functions of obtained states without cross-Kerr nonlinearity are displayed. The clear difference of fidelities shows the strong nonreciprocity of cat state generation in counter-propagating edge cavity modes. When the cross-Kerr nonlinearity is introduced, the changed Wigner functions of obtained states are displayed in Fig.\,\ref{fig4_sm}(c)(d). Here, we chose the cross-Kerr nonlinear strength to be twice as large as the self-Kerr nonlinear strength. This is typically derived from the optomechanical system (or cavity-atom system) where two optical modes are coupled to a common mechanical oscillator (atom). The Wigner functions of obtained states show that the cross-Kerr nonlinearity hardly change the fidelity in CW mode ($F_{\circlearrowright}=0.99$ for both $ \chi_c/\chi = 0 $ and $ \chi_c/\chi = 2 $ cases), but increase the fidelity in CCW mode ($F_{\circlearrowleft}=0.78$ for $\chi_c/\chi = 2$ and $F_{\circlearrowleft}=0.73$ for $\chi_c/\chi = 0$). In Fig.\,\ref{fig4_sm}(e), we demonstrate the fidelities $F$ and nonreciprocal rate $R_F$ against the strength of cross-Kerr nonlinearity. The results also imply that the cross-Kerr nonlinearity will reduce the nonreciprocity of cat state generation mainly by enhancing the quality of obtained state in CCW mode. As some self-Kerr nonlinearity generation methods accompany with cross-Kerr nonlinearity may have potential superiority over other methods such as feasibility in experiment, people can flexibly decide whether to adopt the cross-Kerr nonlinearity according to their tolerance of nonreciprocal rate in realistic situation. For example, if one can accept a nonreciprocal rate that not less than 0.2, then they can adopt the cross-Kerr nonlinearity $ \chi_c/\chi=2 $ caused by the optomechanical coupling or atom-resonator coupling in our scheme. If they can't accept this nonreciprocal rate, they can use other approaches such as Kerr material and electromagnetically induced transparency to generate self-Kerr nonlinearity without generating cross-Kerr nonlinearity\,\cite{Schmidt1996,Brasch2016}.

\section{Additional discussion for nonreciprocal transmission}
\begin{figure}[t]
	\includegraphics[width=12cm]{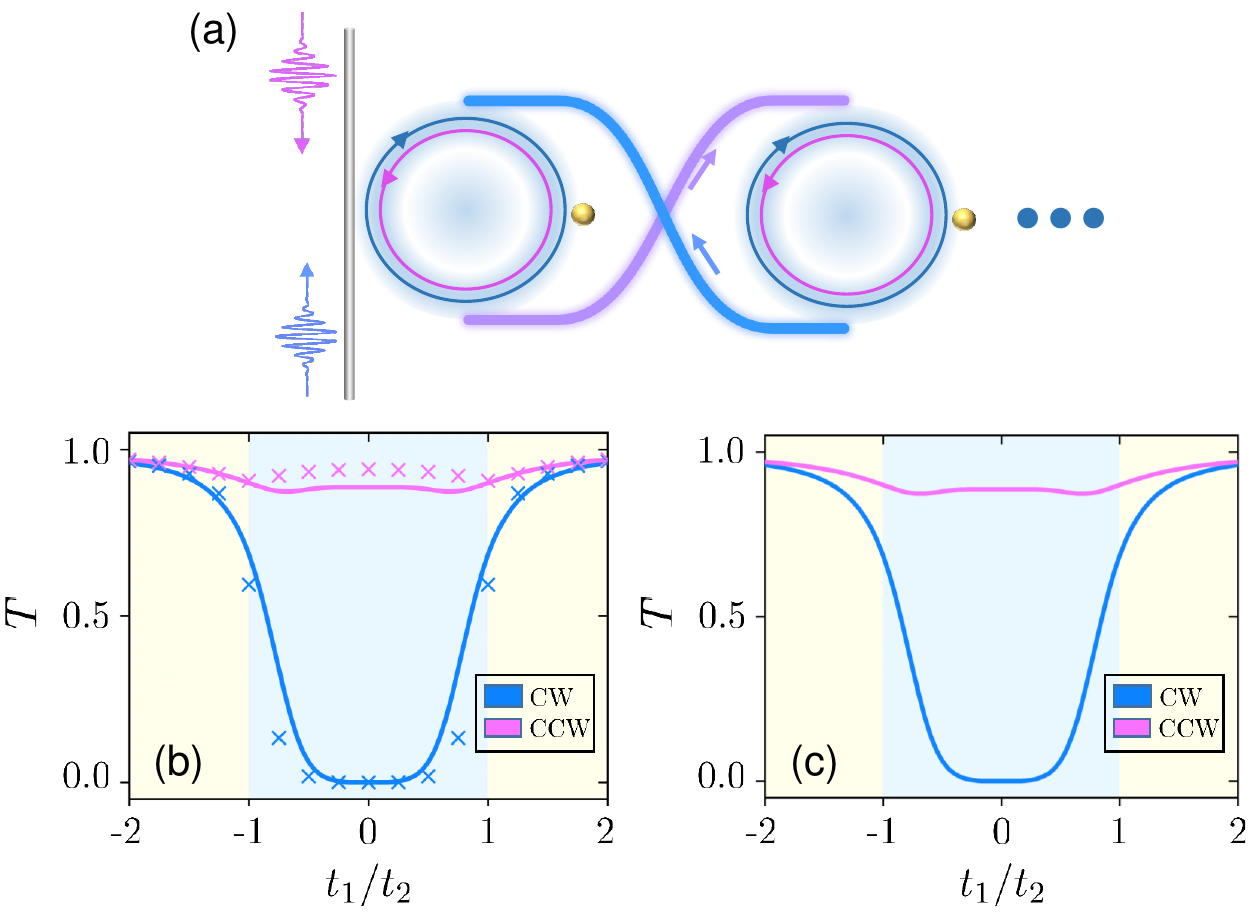}
	\caption{(a) Edge cavity is driven from the opposite direction through an extra fiber coupled on the left side.The transmission rate $ T $ of (b) linear system and (c) system with the presence of Kerr nonlinearity $ \chi = 4\pi\kappa $ as a function of $ t_{1}/t_{2} $. The solid lines denote the numerical results and the crosses in (b) denote the analytical results of linear system. The other parameters are $ \gamma=\kappa $, $ \epsilon=0.08\kappa $, $ N=5 $.}
	\label{fig5_sm}
\end{figure}

In the main text, we have shown the excitation spectrum at the Kerr sideband to demonstrate the chiral Kerr effect. Actually, this chiral effect is related to the nonreciprocal transmission rate of system. As a support for the main text, in this section, we first analytically solve the transmission rate of linear system, and give the numerical verification. Then we give the numerical results of the transimission rate of the system in the presence of Kerr nonlinearity. Through an extra fiber coupled to the left edge cavity [see Fig.\,\ref{fig5_sm}(a)], we drive the cavity from the opposite direction with power $ P_d $ and frequency $ \omega_d $. For the linear system, the total Hamiltonian can be written as $H_{\rm tot}=H_c+\tilde{H}_l+H_d$, where
\begin{align}
	H_{\rm c}=&\sum^{N}_{j=1}[\omega_a (a_{j,\circlearrowright}^{\dagger}a_{j,\circlearrowright}+ a_{j,\circlearrowleft}^{\dagger}a_{j,\circlearrowleft})\!+\!(t_{1} a_{j,\circlearrowright}^{\dagger}a_{j,\circlearrowleft}+h.c.)],\\
	\tilde{H}_{l} =& \sum^{N-1}_{j}  (t_{2} a_{j,\circlearrowleft}^{\dagger}a_{j+1,\circlearrowright}+t_{2} a_{j+1,\circlearrowright}^{\dagger}a_{j,\circlearrowleft}),\\
	H_d=&i\sqrt{2\gamma}(\epsilon e^{-i\omega_d t} a_{1,\circlearrowright/\circlearrowleft}^{\dagger}-h.c.).
\end{align}
Here, $ \gamma $ is the loss induced by the extra driving fiber and $ \epsilon=\sqrt{2\pi P_d/\hbar\omega_d} $. The QLE of system with probe field can be written in the matrix form $ \dot{a} = -Ma+u_{\circlearrowright/\circlearrowleft} $ where $ a =(a_{1,\circlearrowright},a_{1,\circlearrowleft},a_{2,\circlearrowright},a_{2,\circlearrowleft},\cdots)^{T} $, $ u_{\circlearrowright} =(\sqrt{2\gamma}\epsilon,...)^{T} $, $ u_{\circlearrowleft} =(0,\sqrt{2\gamma}\epsilon,...)^{T} $ and
\begin{equation}
	\begin{aligned}
		M &=\left(
		\begin{array}{ccccc}
			\gamma+\kappa+i\Delta_p&it_1&0&0 & \cdots \\ 
			it_1&\gamma+\kappa+i\Delta_p&it_2&0 & \cdots \\
			0 & it_2 & \kappa+i\Delta_p & it_1& \cdots \\
			0&0&it_1&\kappa+i\Delta_p& \cdots  \\
			\vdots & \vdots & \vdots & \vdots & \ddots \\
		\end{array}
		\right)
	\end{aligned}
\end{equation}
is a $ N \times N $ matrix. Here, $ \kappa $ is the loss induced by the coupling between cavity and unidirectional fibers. The steady-state solution of dynamical variables are given by $ \bar{a}_{\circlearrowright/\circlearrowleft} = M^{-1}u_{\circlearrowright/\circlearrowleft} $. 

To obtain the inverse matrix of $ M $, we start from the $ (N-2) \times (N-2) $ matrix without the first two columns and rows, denoted as $ D $. Writing $D$ in block form\,\cite{Bernstein2005}
\begin{equation}
	D=
	\left(
	\begin{array}{c|cccc}
		\kappa+i\Delta_p&it_1&0&0 & \cdots \\ \hline 
		it_1&\kappa+i\Delta_p&it_2&0 & \cdots \\
		0 & it_2 & \kappa+i\Delta_p & it_1& \cdots \\
		0&0&it_1&\kappa+i\Delta_p& \cdots  \\
		\vdots & \vdots & \vdots & \vdots & \ddots \\
	\end{array}  
	\right) = 
	\left(
	\begin{array}{cc}
		E&F \\
		G&H \\
	\end{array}  
	\right),
\end{equation}
the inverse of matrix $ D $ can be written as
\begin{equation} \label{S22}
	\begin{aligned}
		D^{-1}=
		\left(
		\begin{array}{cc}
			(E-FH^{-1}G)^{-1}&\cdots \\
			\vdots&\ddots \\
		\end{array}  
		\right)
		= 
		\left(
		\begin{array}{cc}
			(\kappa+i\Delta_p+t_1^2[H^{-1}]_{11})^{-1}&\cdots \\
			\vdots&\ddots \\
		\end{array}  
		\right),
	\end{aligned}
\end{equation}
where $ [H^{-1}]_{11} $ is the first row and first column in the inverse matrix of $ H $. The inverse of matrix $ H $ can be obtained following the same step
\begin{equation} \label{S23}
	\begin{aligned}
		\setlength{\arraycolsep}{0.2pt}
		H^{-1}&=
		\left(
		\begin{array}{cc}
			(\kappa+i\Delta_p+t_2^2[R^{-1}]_{11})^{-1}&\cdots \\
			\vdots&\ddots \\
		\end{array}
		\right)
	\end{aligned}
\end{equation}
with
\begin{equation}
	R=
	\left(
	\begin{array}{ccccc}
		\kappa+i\Delta_p&it_1&0&0 & \cdots \\
		it_1&\kappa+i\Delta_p&it_2&0 & \cdots \\
		0 & it_2 & \kappa+i\Delta_p & it_1& \cdots \\
		0&0&it_1&\kappa+i\Delta_p& \cdots  \\
		\vdots & \vdots & \vdots & \vdots & \ddots \\
	\end{array}
	\right).
\end{equation}
Here, $ R $ is a $ (N-4) \times (N-4) $ matrix. Comparing to the matrix $ D $, $ R $ has the similar form, except that the first two columns and rows are missing. Suppose that $ N $ is large enough, we have $ [D^{-1}]_{11} \approx [R^{-1}]_{11} $. By combining Eq.\,(\ref{S22}) and Eq.\,(\ref{S23}), we get $ [D^{-1}]_{11} \approx [\kappa+i\Delta_p+t_1^2(\kappa+i\Delta_p+t_2^2[D^{-1}]_{11})^{-1}]^{-1} $ which yields
\begin{eqnarray}
	[D^{-1}]_{11} = \frac{-t_1^2+t_2^2-(\kappa+i\Delta_p)^2 \pm \sqrt{4t_2^2(\kappa+i\Delta_p)^2+(-t_1^2+t_2^2-(\kappa+i\Delta_p)^2)^2}}{2t_2^2(\kappa+i\Delta_p)}.
\end{eqnarray}
Now we write the matrix $ M $ in the form
\begin{equation}
	\begin{aligned}
		M &=
		\left(
		\begin{array}{cc|ccc}
			\gamma+\kappa+i\Delta_p&it_1&0&0 & \cdots \\ 
			it_1&\gamma+\kappa+i\Delta_p&it_2&0 & \cdots \\  \hline
			0 & it_2 & \kappa+i\Delta_p & it_1& \cdots \\
			0&0&it_1&\kappa+i\Delta_p& \cdots  \\
			\vdots & \vdots & \vdots & \vdots & \ddots \\
		\end{array}  
		\right) = 
		\left(
		\begin{array}{cc}
			A&B \\
			C&D \\
		\end{array}  
		\right).
	\end{aligned}
\end{equation}
The inverse of matrix $ M $ can be written as
\begin{equation}
	\begin{aligned}
		\setlength{\arraycolsep}{0.2pt}
		M^{-1}=
		\left(
		\begin{array}{cc}
			(A-BD^{-1}C)^{-1}&\sim \\
			\sim&\sim \\
		\end{array}  
		\right),\\
	\end{aligned}
\end{equation}
where
\begin{equation}
	\begin{aligned}
		\setlength{\arraycolsep}{1.2pt}
		A&=
		\left(
		\begin{array}{cc}
			\gamma+\kappa+i\Delta_p&it_1 \\
			it_1&\gamma+\kappa+i\Delta_p \\
		\end{array}  
		\right),\\
		BD^{-1}C &= 
		\left(
		\begin{array}{cc}
			0&0 \\
			0&-t_2^2[D^{-1}]_{11} \\
		\end{array}  
		\right).
	\end{aligned}
\end{equation}
Then we have
\begin{equation}
	\begin{aligned}
		\left[M^{-1}\right]_{11} = \frac{\gamma+\kappa+i\Delta_p+[D^{-1}]_{11}t_2^2}{t_1^2+(\gamma+\kappa+i\Delta_p)(\gamma+\kappa+i\Delta_p+[D^{-1}]_{11}t_2^2)},\\
		\left[M^{-1}\right]_{22} = \frac{\gamma+\kappa+i\Delta_p}{t_1^2+(\gamma+\kappa+i\Delta_p)(\gamma+\kappa+i\Delta_p+[D^{-1}]_{11}t_2^2)}.
	\end{aligned}
	\label{}
\end{equation}
The steady-state value of $ a_{1,\circlearrowright/\circlearrowleft} $ is
\begin{eqnarray}
	\bar{a}_{1,\circlearrowright} &= \left[M^{-1}\right]_{11}\sqrt{2\gamma}\epsilon,\\
	\bar{a}_{1,\circlearrowleft} &= \left[M^{-1}\right]_{22}\sqrt{2\gamma}\epsilon.
\end{eqnarray}
For $ \gamma=\kappa $, we derive the transmission rate of input field from $T_{\circlearrowright/\circlearrowleft} = |1-\frac{\sqrt{2\gamma}}{\epsilon}\bar{a}_{1,\circlearrowright/\circlearrowleft}|^2 = |t_{\circlearrowright/\circlearrowleft}|^2$ and obtain
\begin{eqnarray}
	t_{\circlearrowright}&=\frac{(\Delta_p-2i\kappa)t_1^2+\Delta_p(t_2^2-(\Delta_p-i\kappa)(\Delta_p-3i\kappa)+A)}{\Delta_p t_1^2+(\Delta_p-2i\kappa)(t_2^2-(\Delta_p-i\kappa)(\Delta_p-3i\kappa)+A)},\\
	t_{\circlearrowleft}&=\frac{\Delta_p t_1^2+(\Delta_p-2\kappa)(t_2^2-(\Delta_p-i\kappa)(\Delta_p+i\kappa)+A)}{\Delta_p  t_1^2+(\Delta_p-2\kappa)(t_2^2-(\Delta_p-i\kappa)(\Delta_p-3i\kappa)+A)},
\end{eqnarray}
with
\begin{eqnarray}
	A=\sqrt{(t_1^2-(\Delta_p-t_2-i\kappa)^2)(t_1^2-(\Delta_p+t_2-i\kappa)^2)}.
\end{eqnarray}
In Fig.\,\ref{fig5_sm}(b), we show the transmission rate of linear system for zero detuning as a function of $ t_1/t_2 $. In topological phase ($t_1<t_2$), the transmission rate of CW mode almost reach $ 0 $ for a wide range, whereas the transmission rate of CCW remain high, showing the obvious nonreciprocal transmission rate of the edge cavity. This nonreciprocity reduce significantly when the system enters into trivial phase ($ t_1>t_2 $), as the transmission rate of CW mode increases drastically and becomes closer to the transmission rate of CCW mode. The analytical results are roughly in agreement with numerical results. The mismatch comes from: $ (1) $ the finite array length in numerical process and assumptive infinite array length in analytical process and $ (2) $ the missing of jump term in analytical QLE. In the presence of Kerr nonlinearity, we plot the numerical results of transmission rate for zero detuning in Fig.\,\ref{fig5_sm}(c). Comparing to the chiral Kerr excitation showed in main text, the change of nonreciprocal transmission in topological phase and trivial phase is even more obvious. The nonreciprocal transmission of optical field in topological phase also originates from the chirality mode occupations in the edge cavity, it has wide applications in the engineering of optical devices such as optical isolators, circulators and nonreciprocal single-photon sources, as well as integrated optics and quantum network~\cite{Caloz2018,Lodahl2017,Jalas2013}.

\section{Discussion of experimental feasibility}

As a support for the main text, here we discuss the experimental feasibility of the chosen key system parameters in our proposal. In a microresonator, the Kerr nonlinear parameter is $\chi = \hbar \omega_a^2 c n_2 / (n^2 V_{\rm eff})$, where $n_2(n)$ is the nonlinear(linear) refractive index, $c$ is the speed of light in vacuum,  $V_{\rm eff}$ is the nonlinear optical mode\,\cite{Palomo2017}. According to the current accessible experimental conditions, $V_{\rm eff}$ is typically $10–10^3 \mu \rm m^3$\,\cite{Vahala2003,Spillane2005}, $ n_2 \sim 10^{-14} $ can be realized in device constructed from potassium titanyl phosphate crystal\,\cite{Zielinska2017}. With the resonator frequency $ \omega_a \sim 1 $PHz and the quality factor $ Q \sim 6 \times 10^8 $\,\cite{Armani2003,Ostby2009}, the dissipation rate $ \kappa_f = \omega_a/Q \sim 1.6 $MHz and strong kerr nonlinearity for the microresonator $ \chi \approx 3 \kappa_f $ are feasible to achieve. The backscattering coupling can be experimentally tuned by changing the distance between the scatterer and resonator\,\cite{Zhu2010,Peng2016} and reach $ t_1 \approx 1.5 \kappa_f $. As derived in Sec.\,\ref{sectionI}, the off-site coupling is $ t_2 = 2\kappa_f $. Moreover, we note the Kerr nonlinearity can be enhanced by various approaches\,\cite{Lu2013, Bartkowiak2014, Lu2015, Higo2018}. Meanwhile, the coupling-loss relation and nonlinearity-loss relation can be achieved by changing the total loss rate $ \kappa$. For example, in the resonator made from $ \rm Er^{3+}$-doped silica, a gain rate $\mathcal{G}$ can be introduced by the optically pumping $ \rm Er^{3+} $ ions with a pump laser\,\cite{Peng2014}. For gain rate chosing around $ \mathcal{G}/\kappa_f \in [0,0.9]$, the total loss rate can be reduced to $ \kappa/\kappa_f=(\kappa_f-\mathcal{G})/\kappa_f\in [0.1,1]$, thus the coupling-loss relations $t_1/\kappa\in[0,15]$, $t_2/\kappa\in[2,20]$ and nonlinearity-loss relation $\chi/\kappa\in[3,30]$ in our proposal could be achieved with feasible experimental parameters.

\end{document}